\documentclass[a4paper,aps,pra,twocolumn,preprintnumbers,amsmath,amssymb]{revtex4}
\usepackage{bbm}
\usepackage{amsmath}
\usepackage{verbatim}
\usepackage{graphicx}
\bibliography{References}
\begin{document}

\title{Quantum Teleportation of Dynamics\\ and Effective Interactions between Remote Systems}

\author{Christine A. Muschik$^{1}$, Klemens Hammerer$^{2}$, Eugene S. Polzik$^{3}$, and Ignacio J. Cirac$^{4}$}

\affiliation{
$^{1}$ ICFO-Institut de Ci\`{e}ncies Fot\`{o}niques,
Mediterranean
Technology Park, 08860 Castelldefels (Barcelona), Spain.\\
$^{2}$ Institute for Theoretical Physics, Institute for Gravitational Physics (Albert Einstein Institute), Leibniz University Hannover, Callinstr. 38, 30167 Hannover, Germany.\\
$^{3}$ Niels Bohr Institute, Danish Quantum Optics Center QUANTOP,
Copenhagen University, Blegdamsvej 17, 2100 Copenhagen, Denmark.\\
$^{4}$ Max-Planck-Institut f\"ur Quantenoptik,
Hans-Kopfermann-Str. 1, 85748 Garching, Germany.}

\begin{abstract}
Most protocols for quantum information processing consist of a series of quantum gates, which are applied sequentially. In contrast, interactions between matter and fields, for example, as well as measurements such as homodyne detection of light, are typically continuous in time. We show how the ability to perform quantum operations continuously and deterministically can be leveraged for inducing nonlocal dynamics between two separate parties. We introduce a scheme for the engineering of an interaction between two remote systems and present a protocol that induces a dynamics in one of the parties, that is controlled by the other one. Both schemes apply to continuous variable systems, run continuously in time and are based on real-time feedback.
\end{abstract}


\maketitle 

Most protocols in quantum information science (QIS) are discrete in the sense that they consist of a sequence of unitary operations and measurements. Schemes for quantum teleportation or dense coding are typical examples. These elementary protocols are the building blocks of other applications such as quantum repeaters or quantum computing.
However, some implementations are intrinsically continuous. The most prominent example are atomic ensembles interacting with light, where schemes based on the continuous detection of quadrature operators are realized~\cite{BvL05,HSP10,Review2}.
In this system, protocols can be performed that are intrinsically deterministic and continuous in time.
Here we address the question how this property can be exploited by designing primitives that take advantage of this fact.
Continuous schemes have been devised in several subfields of QIS, for example for phase estimation~\cite{Wiseman95,Armen2002}, error correction~\cite{Kerckhoff2011}, and the preparation and protection of quantum states~\cite{Bose2004,Mancini2005,Vu2012,Xue2013}, in particular in the context of dissipative schemes~\cite{Clark2003,Parkins2006,Diehl2008,Sorensen2011,Krauter2011,Muschik2011,Lukin2013}.
Here we introduce two protocols that achieve a qualitatively new goal - to control and transmit quantum evolutions between remote locations. We consider two separate systems that cannot interact directly but exchange quantum states and classical information. One scheme implements an effective nonlocal dynamics, where the two systems evolve as if they were interacting with each other. The other protocol realizes the quantum teleportation of a time evolution, which uses the dynamics of one system to steer the evolution of the other.

\begin{figure}
 \includegraphics[width=\columnwidth]{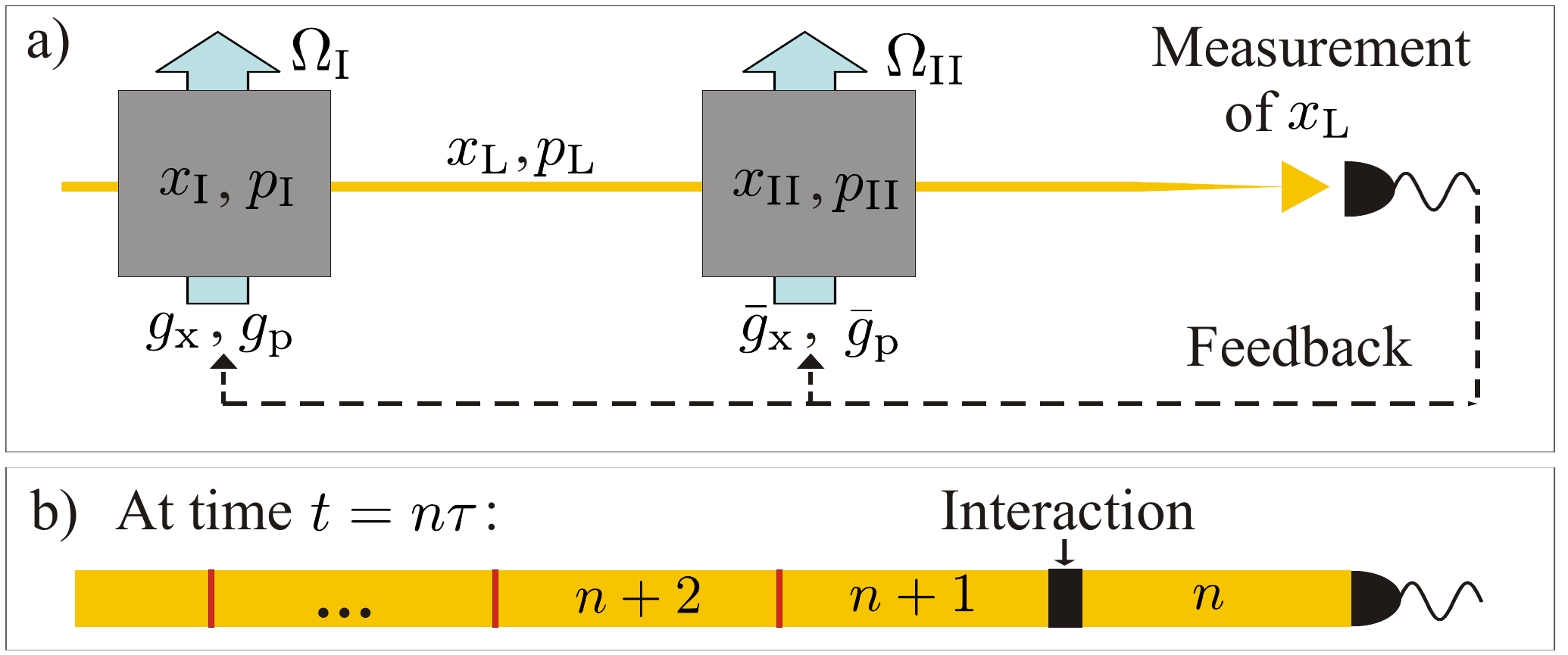}
 \caption{Dynamical teleportation and creation of an interaction between two remote systems. a) The setup consists of two atomic ensembles in constant magnetic fields. A freely propagating light field interacts with both samples and is continuously measured. The result is fed back to the atoms. b) Illustration of the light-matter interaction in terms of discretized spatially localized light modes.}
 \label{Fig:Setup}
 \end{figure}
 Fig.~\ref{Fig:Setup}a shows the setup under consideration. Two spin ensembles interact with a propagating light field, which is constantly measured. By performing real-time feedback on both samples, an effective interaction is established between the systems. Since this is done continuously, the dynamics of the two systems corresponds to the evolution under the desired interaction Hamiltonian at any instant of time. Remarkably, this scheme results ideally in a joint unitary evolution of the two remote systems. This is the case even though the protocol is based on measurements yielding random outcomes and therefore random projections of the states involved~\cite{VC09}. We show that using a quantum nondemolition (QND) interaction between spins and light, any Hamiltonian that is quadratic in the atomic operators can be realized by tuning the feedback operation only, i.e. without variation of the system parameters. In the ideal case, any quadratic Hamiltonian can be implemented perfectly.

The second protocol realizes a continuous teleportation. Teleportation~\cite{BeBCJPW93} offers a practical solution to the delicate task of transmitting quantum states~\cite{FootnoteTeleportation}. It is a prerequisite for quantum networks~\cite{Briegel98,Kimble2008} and a primitive for quantum computation~\cite{Gottesmann99}. A standard teleportation scheme consists of three separate steps, which involve (i) establishment of a highly entangled link (ii) a projective measurement that destroys the state to be teleported, and (iii) the recovery of the state on the receiver's side by applying a feedback operation. Here, we consider a continuous process involving weakly entangled states and measurements that disturb the quantum state only a little at each instant of time. In contrast to previous approaches, which transmit a (static) state using a single feedback operation at the end of the protocol, we consider the transmission of a whole time evolution using real-time feedback. To illustrate this point, we consider the continuous teleportation between two parties, Bob and Charlie, when a time-dependent magnetic field is applied to Charlie's system. The resulting displacements on Charlie's side translate into a corresponding evolution on Bob's system, which evolves as if it was placed in a time-dependent magnetic field: it evolves as if interacting with a field whose time dependence is determined by Charlie's evolution.

We consider two separate systems that are characterized in terms
of continuous variables $x_{\text{I}}$,  $p_{\text{I}}$ and
$x_{\text{II}}$, $p_{\text{II}}$, which commute canonically
$[x,p]=i$. We assume that both systems can be rotated locally and
interact with a propagating auxiliary bosonic system via a QND
interaction (Eq.~(\ref{Eq:QND})). For concreteness, we
consider two atomic spin ensembles interacting with coherent light~\cite{FootnoteEnsembles}.
The free evolution of the atomic system
\begin{eqnarray*}
H_{\text{A}}&=&\frac{\Omega_{\text{I}}}{2}\left(x_{\text{I}}^2+p_{\text{I}}^2\right)+\frac{\Omega_{\text{II}}}{2}\left(x_{\text{II}}^2+p_{\text{II}}^2\right),
\end{eqnarray*}
describes the atomic rotation (Larmor precession) in homogeneous
magnetic fields along $\hat{x}$ with Larmor frequencies
$\Omega_{\text{I}}$ and $\Omega_{\text{II}}$ (see
Fig.~\ref{Fig:Setup}a). In the following, we use transformed atomic variables
\begin{eqnarray}\label{Eq:RotationMatrix}
\left(
  \begin{array}{c}
   \! \tilde{x}_{\text{I/II}}\! \\
   \! \tilde{p}_{\text{I/II}}\! \\
  \end{array}
\right)\!\!&=&\!\!R(\Omega_{\text{\tiny{I/II}}})\left(
  \begin{array}{c}
    x_{\text{I/II}} \\
    p_{\text{I/II}} \\
  \end{array}
\right),\nonumber \\
R(\Omega_{\text{\tiny{I/II}}}) \!\!&=& \!\!\left(
                    \begin{array}{cc}
                     \! \cos(\Omega_{\text{I/II}}t) & -\sin(\Omega_{\text{I/II}}t)\! \\
                      \!\sin(\Omega_{\text{I/II}}t)  & \cos(\Omega_{\text{I/II}}t)\!  \\
                    \end{array}
                  \right)\!\left(
  \begin{array}{c}
    \!x_{\text{I/II}}\! \\
   \! p_{\text{I/II}}\! \\
  \end{array}
\right),
\end{eqnarray}
which rotate at the Larmor frequencies of the respective fields.
The light propagates along $\hat{z}$, passing both ensembles. We adopt here a discretized one-dimensional model. The interaction time $T$ is divided into $N$ infinitesimally small time steps of length
$\tau$ and the light is accordingly described in
terms of $N$ short pulse pieces. The quadratures associated with the $\text{n}^{\text{th}}$ localized
light mode are denoted by $x_{\text{L,n}}$ and $p_{\text{L,n}}$
with $[x_{\text{L,n}},p_{\text{L,n'}}]=i\delta_{\text{\tiny{n,n'}}}$.
The $\text{n}^{\text{th}}$ pulse piece interacts
with the atoms during the time window from $t=(n-1)\tau$ to $t=n\tau$
(Fig.~\ref{Fig:Setup}b) according to the interaction
Hamiltonian~\cite{FootnoteNeglectR}
\begin{eqnarray}\label{Eq:QND}
H_{\text{\tiny{QND,n}}}&=&\frac{\kappa}{\sqrt{N}\tau}\left(p_{\text{I}}(n\tau)+p_{\text{II}}(n\tau)\right)
p_{\text{L,n}},
\end{eqnarray}
where $\kappa$ is a dimensionless coupling constant. Such a QND interaction can, for example,
be realized using a Faraday interaction in atomic
vapors~\cite{DCZ00,JKP01} or in optomechanical systems~\cite{Optomechanics}. The total Hamiltonian is given by
$H\!=\!H_{\text{L}}\!+\!H_{\text{A}}\!+\!\sum_{n=1}^N H_{\text{\tiny{QND,n}}}$, where
$H_{\text{L}}$ accounts for the free propagation of the light.
By virtue of the consecutive QND interactions shown in Fig.~\ref{Fig:Setup}a, the
quantum state of the atoms is mapped to the $x$-quadrature of the
light field. The corresponding input-output relation for an
infinitesimal time step is given by
\begin{eqnarray}\label{Eq:LightField}
x_{\text{{L,n}}}^{\text{{out}}}\!\!&=&\!\!x_{\text{{L,n}}}^{\text{{in}}}+\frac{\kappa}{\sqrt{N}}\ \!\vec{V}_{\text{L}}^{T}(n \tau)\vec{R}_{\text{A}}([n-1]\tau),
\end{eqnarray}
with
\begin{eqnarray*}
\vec{V}_{\text{L}}^{T}(t)\!\!&=&\!\!\left( \!-\sin(\Omega_I t) , \cos(\Omega_I t) ,  \!-\sin(\Omega_{II} t) , \cos(\Omega_{II} t) \right),\\
\vec{R}_{\text{A}}^T(t)&=&\left(\tilde{x}_{I}(t),\tilde{p}_{I}(t),\tilde{x}_{II}(t),\tilde{p}_{II}(t)\right),
\end{eqnarray*}
where $\Omega \tau\ll 1$ has been assumed~\cite{FootnoteRotation}.
The $p$-quadrature is conserved
$p_{\text{L,n}}^{\text{out}}=p_{\text{L,n}}^{\text{in}}$. The atomic evolution is given by
\begin{eqnarray*}
\left(%
\begin{array}{c}
  \!\!\tilde{x}_\text{\tiny{I/II}}(n\tau)\!\!\\
 \!\! \tilde{p}_{\text{\tiny{I/II}}}(n\tau)\!\!\\
\end{array}%
\right)\!\!&=&\!\!\left(%
\begin{array}{c}
  \!\!\tilde{x}_\text{\tiny{I/II}}([n\!-\!1]\tau)\!\!\\
  \!\!\tilde{p}_{\text{\tiny{I/II}}}([n\!-\!1]\tau)\!\!\\
\end{array}%
\right)\!+\!\frac{\kappa}{\sqrt{N}}\!\left(%
\begin{array}{c}
  \!\!\cos(\Omega_{\text{\tiny{I/II}}} n\tau) \!\!\\
  \!\!\sin(\Omega_{\text{\tiny{I/II}}} n\tau) \!\!\\
\end{array}%
\right)\!p_{\text{L,n}}^{\text{in}}.
\end{eqnarray*}
We assume that $x_{\text{L}}$ is continuously measured and that the result is
instantaneously fed back onto the atoms by applying a conditional displacement. We apply here feedback operations with temporally modulated gain
factors $\frac{1}{\sqrt{N}}\ \!g_{\text\tiny{{x,I/II}}}(t)$ and
$\frac{1}{\sqrt{N}}\ \! g_{\text\tiny{{p,I/II}}}(t)$ such that
\begin{eqnarray}\label{Eq:Feedback}
\left(%
\begin{array}{c}
  \!\!\tilde{x}_{\text{\tiny{I/II}}}^{\text{fin}}(n\tau)\!\! \\
  \!\!\tilde{p}_{\text{\tiny{I/II}}}^{\text{fin}}(n\tau)\!\!\\
\end{array}%
\right)\!\!\!&=&\!\!\!\left(%
\begin{array}{c}
  \!\!\tilde{x}_\text{\tiny{I/II}}(n\tau)\!\!\\
  \!\!\tilde{p}_{\text{\tiny{I/II}}}(n\tau)\!\!\\
\end{array}%
\right)\!+\!\frac{1}{\sqrt{N}}\!\left(%
\begin{array}{c}
\!\!\! g_{\text\tiny{{x,I/II}}}(n\tau)\!\!\! \\
\!\!\! g_{\text\tiny{{p,I/II}}}(n\tau)\!\! \!\\
\end{array}%
\right)\!x_{\text{L,n}}^{\text{out}}.
\end{eqnarray}
\\We show now how this setup can be used to establish an arbitrary quadratic interaction between two ensembles. Using suitable local rotations~\cite{KrHGC03}, any interaction Hamiltonian for two continuous variable systems, which is quadratic in the system operators, can be expressed as
\begin{eqnarray}\label{Eq:quadraticHamiltonian}
H=\gamma\left(\mu
H_{\text{A}}+\nu H_{\text{P}}\right)=\gamma\left(Z x_{1}p_{2}+\frac{1}{Z}p_1x_2\right),
\end{eqnarray}
where $\mu=\frac{1}{2}\left(Z+\frac{1}{Z}\right)$ and
$\nu=\frac{1}{2}\left(Z-\frac{1}{Z}\right)$~\cite{FootnoteHamiltonian1}. $\gamma$ characterizes the
overall coupling strength of the interaction while $Z$ parametrizes the imbalance between the active (entanglement creating) and the passive (energy conserving) components, $H_{\text{A}}$ and $ H_{\text{P}}$~\cite{FootnoteHamiltonian2}.
After the light-matter interactions, the measured quadrature $x_{\text{L}}$ contains information on both ensembles (see Eq.~(\ref{Eq:LightField})). Feedback of $x_{\text{L}}$ according to Eq.~(\ref{Eq:Feedback}), leads therefore to terms which correspond to the evolution under both, local and interaction Hamiltonians. The feedback of information of each ensemble onto itself leads to an effective evolution according to local squeezing Hamiltonians.
In order to suppress these contributions, we use the fact that the quadratures of each ensemble are mapped to the light with an oscillatory time dependence.
For implementing a purely nonlocal evolution, the gain function for ensemble I (II) is chosen to oscillate with $\Omega_{II}$ ($\Omega_{I}$), such that information is transferred with high efficiency between the ensembles, while contributions due to the feedback from the samples to themselves are out of phase and average out.
This becomes apparent by considering the time varying gain functions
$g_{\text\tiny{{x,I/II}}}(t)=\frac{1}{\sqrt{N}}g_{\text{\tiny{a/b}}}\sin(\Omega_{\text{\tiny{a/b}}} t)$ and $g_{\text\tiny{{p,I/II}}}(t)=\frac{1}{\sqrt{N}}g_{\text{\tiny{b/a}}}\cos(\Omega_{\text{a/b}} t)$. In the continuous limit,
\begin{eqnarray*}
\left(%
\begin{array}{c}
 \! \!\!\dot{\tilde{x}}_{\text{I}}^{\text{fin}}\!(t)\!\!\! \\
 \! \!\!\dot{\tilde{p}}_{\text{I}}^{\text{fin}}\! (t)\!\!\!\\
\end{array}%
\right)\!\!\!&=&\!\!\!\frac{\kappa}{T}\!\!\left[\!\!M_{g_{\text{a}},g_{\text{\tiny{b}}}}^{\Omega_{\text{a}},\!\Omega_{\text{I}}}\!(t)\!\!\left(%
\begin{array}{c}
\! \!\! \tilde{x}_{\text{I}}(t)\!\!\!\! \\
 \!\!\! \tilde{p}_{\text{I}}(t)\!\! \\
\end{array}%
\right)\!\!+\!
M_{g_{\text{a}},g_{\text{\tiny{b}}}}^{\Omega_{\text{a}}\!,\Omega_{\text{II}}}\!(t)\!\!\left(%
\begin{array}{c}
\! \!\! \tilde{x}_{\text{II}}(t)\!\!\! \\
 \!\!\! \tilde{p}_{\text{II}}(t)\!\!\! \\
\end{array}%
\right)\!\!\right]
\!\!+\!\!\left(%
\begin{array}{c}
  \!\!\!\mathcal{N}_{\text{\tiny{x,\!I}}}\!(t) \!\!\! \\
 \! \!\!\mathcal{N}_{\text{\tiny{p,\!I}}}\!(t) \!\! \!\\
\end{array}%
\right)\!,\\
\left(%
\begin{array}{c}
 \! \!\!\dot{\tilde{x}}_{\text{II}}^{\text{fin}}\!(t)\!\!\! \\
 \! \!\!\dot{\tilde{p}}_{\text{II}}^{\text{fin}}\!(t)\!\!\!\\
\end{array}%
\right)\!\!\!&=&\!\!\!\frac{\kappa}{T}\!\!\left[\!\!M_{g_{\text{b}},g_{\text{\tiny{a}}}}^{\Omega_{\text{b}},\!\Omega_{\text{II}}}\!(t)\!\!\left(%
\begin{array}{c}
\! \!\! \tilde{x}_{\text{II}}(t)\!\!\!\! \\
 \!\!\! \tilde{p}_{\text{II}}(t)\!\! \\
\end{array}%
\right)\!\!+\!
M_{g_{\text{b}},g_{\text{\tiny{a}}}}^{\Omega_{\text{b}},\!\Omega_{\text{I}}}\!(t)\!\!\left(%
\begin{array}{c}
\! \!\! \tilde{x}_{\text{I}}(t)\!\!\! \\
 \!\!\! \tilde{p}_{\text{I}}(t)\!\!\! \\
\end{array}%
\right)\!\!\right]
\!\!+\!\!\left(%
\begin{array}{c}
  \!\!\!\mathcal{N}_{\text{\tiny{x,\!II}}}\!(t) \!\!\! \\
 \! \!\!\mathcal{N}_{\text{\tiny{p,\!II}}}\!(t) \!\! \!\\
\end{array}%
\right)\!,
\end{eqnarray*}
where the coupling matrix $M_{g_{1}, g_{2}}^{\Omega_{\text{1}},\Omega_{\text{2}}}(t)$ is defined by
\begin{eqnarray*}
M_{g_{1}, g_{2}}^{\Omega_{1},\Omega_{2}}(t) \!\!= \!\!\left(%
\begin{array}{cc}
   \!\!-g_1 \sin(\Omega_{\text{1}}t)\! \sin(\Omega_{\text{2}}t) &  g_1 \sin(\Omega_{\text{1}}t)\!\cos(\Omega_{\text{2}}t) \!\! \\
  \!\! -g_2 \cos(\Omega_{\text{1}}t)\!\sin(\Omega_{\text{2}}t)&  g_2 \cos(\Omega_{\text{1}}t)\!\cos(\Omega_{\text{2}}t) \!\! \\
\end{array}%
\right)\!.
\end{eqnarray*}
$\mathcal{N}_{\text{\tiny{x,I/II}}}$ and $\mathcal{N}_{\text{\tiny{p,I/II}}}$ are noise terms due to the mapping of the input light field onto the atomic systems,
\begin{eqnarray*}
\left(%
\begin{array}{c}
  \! \!\!\mathcal{N}_{\text{\tiny{x,I/II}}}(t) \!\!\! \\
  \! \!\!\mathcal{N}_{\text{\tiny{p,I/II}}}(t) \!\!\!\\
\end{array}%
\right) \!\!\!&=& \!\!\!\frac{1}{\sqrt{T}} \!\!
\left(
  \begin{array}{cc}
     \!\!g_{\text{\tiny{a/b}}}\!\sin(\Omega_{\text{\tiny{a\!/\!b}}}t) & \kappa\! \cos(\Omega_{\text{\tiny{I\!/\!II}}}t) \!\! \\
     \!\!g_{\text{\tiny{b/a}}}\! \cos(\Omega_{\text{\tiny{a\!/\!b}}}t) & \kappa\!\sin(\Omega_{\text{\tiny{I\!/\!II}}}t)  \!\!\\
  \end{array}
\right) \!\!\left(
         \begin{array}{c}
            \!\!\bar{x}_{\text{L}}(ct,\!0) \!\! \\
           \!\! \bar{p}_{\text{L}}(ct,\!0) \!\! \\
         \end{array}
       \right)\!.
\end{eqnarray*}
We use here continuous light modes with quadratures $\bar{x}_{\text{L}}(ct,0)=x_{\text{L,n}}^{\text{in}}/\sqrt{\tau}$ and $\bar{p}_{\text{L}}(ct,0)=p_{\text{L,n}}^{\text{in}}/\sqrt{\tau}$~\cite{FootnoteLightmodes}.
As shown in the Supplemental Material (SM), the equations above can be approximated by their coarse-grained version using coarse-graining time intervals $\Delta t\gg \Omega_{I/II}^{-1}$, $|\Omega_{I}-\Omega_{II}|^{-1}$. In this limit, the coupling matrices $M_{g_{1}, g_{2}}^{\Omega_{1},\Omega_{2}}$ lead to a negligible contribution for $\Omega_{\text{1}}\neq\Omega_{\text{2}}$, since their matrix elements average out. Similarly, they can be approximated by a constant diagonal matrix for $\Omega_{\text{1}}=\Omega_{\text{2}}$~\cite{SupplementalMaterial}.
The noise terms $\mathcal{N}_{\text{\tiny{x,I/II}}}(t)$ and $\mathcal{N}_{\text{\tiny{p,I/II}}}(t)$ give rise to noise modes, which are approximately independent for $\Delta t\gg\Omega_{I/II}^{-1},|\Omega_{I}-\Omega_{II}|^{-1}$, and can therefore be squeezed simultaneously such that their contributions become negligible. A detailed analysis is provided in~\cite{SupplementalMaterial}.
For establishing an interaction according to Eq.~(\ref{Eq:quadraticHamiltonian}) with $\gamma=\frac{\kappa g}{2T}$, we consider the case $\Omega_{\text{a/b}}=\Omega_{\text{II/I}}$, $g_{\text{a/b}}=- g  Z^{\mp1}$, which leads to
\begin{eqnarray*}
\left(
  \begin{array}{c}
    \dot{\tilde{x}}_{\text{I}} \\
    \dot{\tilde{p}}_{\text{I}} \\
  \end{array}
\right)&=&\frac{\kappa g}{2T}\left(
                  \begin{array}{c}
                    \frac{1}{Z}\ \!\tilde{x}_{\text{II}} \\
                    -Z \ \!\tilde{p}_{\text{II}} \\
                  \end{array}
                \right),\ \ \ \
    \left(
  \begin{array}{c}
    \dot{\tilde{x}}_{\text{II}} \\
    \dot{\tilde{p}}_{\text{II}} \\
  \end{array}
\right)=\frac{\kappa g}{2T}\left(
                  \begin{array}{c}
                   Z\ \!\tilde{x}_{\text{I}} \\
                    - \frac{1}{Z} \ \!\tilde{p}_{\text{I}} \\
                  \end{array}
                \right).
\end{eqnarray*}
By tuning the feedback parameters $g_{\text{a/b}}$, it is therefore possible to realize any time evolution that corresponds to a quadratic interaction Hamiltonian.\\
Using a modified configuration, a continuous teleportation can be realized. A teleportation scheme~\cite{BeBCJPW93} involves three parties, - Alice, Bob and Charlie. It allows Alice to teleport an unknown quantum state provided by Charlie to Bob. Here, Charlie's state is
stored in ensemble \!II and teleported to ensemble \!I, representing Bob, while the light field plays the role of Alice. Step (i) in the standard protocol outlined in the introduction corresponds to the interaction between the light and ensemble \!I
resulting in an entangled state. The distribution of entanglement between the remote sites is realized by
the free propagation of the light. Step (ii) corresponds to the interaction of the light with ensemble \!II and the measurement of $x_{\text{L}}$. Step (iii) is implemented in the form of a feedback
operation realizing a conditional displacement on ensemble I, which can be done using magnetic fields.
\begin{figure}
 \includegraphics[width=\columnwidth]{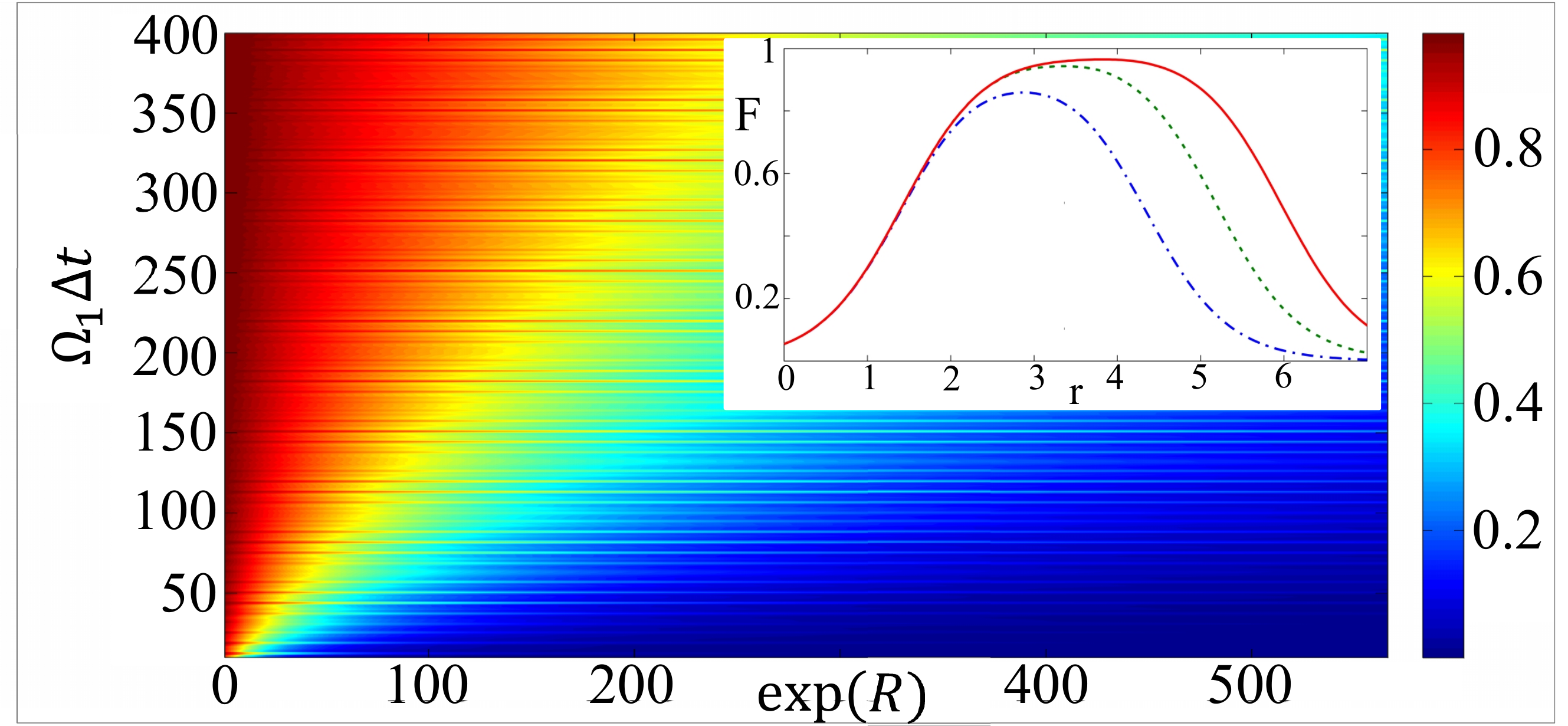}
 \caption{Fidelity for realizing $H_{\text{A}}$ for $\Omega_2=2\Omega_1$ and $g\!=\!\kappa\!=\!1$. The results for implementing $ H_{\text{P}}$ are provided in~\cite{SupplementalMaterial}. Any evolution under a quadratic Hamiltonian can be realized by combining these two interactions with local rotations. ~\cite{KrHGC03,Bennett2002}. The main panel displays the fidelity for optimal squeezing $r_{\text{opt}}$ versus $e^R$ and $\Omega_1\Delta t$ . The inset shows the fidelity versus the squeezing of the light for $R=3$.  The three curves correspond to $\Omega_1\Delta t=100$ (dash-dotted line), $\Omega_1\Delta t=300$ (dashed line) and $\Omega_1\Delta t=500$ (solid line). }
 \label{Fig:Plot1}
 \end{figure}
Using this protocol, the deterministic teleportation of a quantum state between two ensembles has been demonstrated recently~\cite{Krauter2013}. We present now an extension to a time-continuous operation, which facilitates the teleportation of quantum dynamics. For this purpose, we consider the special case of the scheme above where $\Omega_{\text{\tiny{I}}}\!=\!\Omega_{\text{\tiny{II}}}\!=\!\Omega$ and feedback is applied only to the first ensemble $g_{\text\tiny{{x,I}}}(t)=\frac{1}{\sqrt{N}}\bar{g}_{\text{\tiny{x}}}\sin(\Omega t)$, $g_{\text\tiny{{p,I}}}(t)=\frac{1}{\sqrt{N}}\bar{g}_{\text{\tiny{p}}}\cos(\Omega t)$. With this choice,
Eq.~(\ref{Eq:Feedback}) yields
\begin{eqnarray}\label{Eq:Teleportation}
\left(%
\begin{array}{c}
 \!\!\dot{\tilde{x}}_{\text{I}}^{\text{fin}}(t)\!\!\\
 \!\!\dot{\tilde{p}}_{\text{I}}^{\text{fin}}(t)\!\!\\
\end{array}%
\right)\!\!\!&=&\!\!\frac{\kappa}{T}M^{\Omega,\Omega}_{\bar{g}_{\text{\tiny{x}}},\bar{g}_{\text{\tiny{p}}}}\!(t)\!\left(%
\begin{array}{c}
  \!\tilde{x}_{\text{I}}(t)\! \\
  \!\tilde{p}_{\text{I}}(t)\!\\
\end{array}%
\right)\!+\!\frac{\kappa}{T}M^{\Omega,\Omega}_{\bar{g}_{\text{\tiny{x}}},\bar{g}_{\text{\tiny{p}}}}\!(t)\!\left(%
\begin{array}{c}
  \!\tilde{x}_{\text{II}}(t)\! \\
  \!\tilde{p}_{\text{II}}(t)\!\\
\end{array}%
\right)\nonumber\\
\!\!\!&+&\!\!\!\frac{1}{\sqrt{T}}\!\left(
             \begin{array}{cc}
              \!\!\!\bar{g}_{\text{\tiny{x}}} \sin(\Omega t)\! &\! \kappa\cos(\Omega t)\!\! \\
             \!\!\!\bar{g}_{\text{\tiny{p}}} \cos(\Omega t)\! & \!\kappa \sin(\Omega t) \!\!\\
             \end{array}
           \right)\!\!\left(
                   \begin{array}{c}
                    \!\! \bar{x}_{\text{L}}(ct,0) \!\!\\
                   \!\! \bar{p}_{\text{L}}(ct,0)\! \!\\
                   \end{array}
              \right)\!\!.
\end{eqnarray}
The second term in Eq.~(\ref{Eq:Teleportation}) allows us to control Bob's dynamics using Charlie's system.
For demonstrating the transmission of a time evolution, we add an extra Hamiltonian corresponding to a time-dependent
transverse magnetic field on Charlie's side
$H_{\text{extra,II}}=\alpha_{x}(t) x_{\text{II}}+\alpha_p(t)
p_{\text{II}}$,
where $\alpha_{x/p}(t)$ are real time-dependent
functions. Accordingly,
\begin{eqnarray}\label{Eq:EvolutionCharlie}
\left(%
\begin{array}{c}
  \!\!\tilde{x}_{\text{II}}(t)\!\! \\
  \!\!\tilde{p}_{\text{II}}(t)\!\!\\
\end{array}%
\right)
\!\!&=&\int_{0}^t \!\!dt'\left(
                                               \begin{array}{c}
                                                  \!\!\tilde{\alpha}_p(t')  \!\!\\
                                                 \!\! -\tilde{\alpha}_x(t')  \!\!\\
                                               \end{array}
                                             \right)+\left(%
\begin{array}{c}
  \!\!\tilde{x}_{\text{II}}^{\text{in}}\!\! \\
  \!\!\tilde{p}_{\text{II}}^{\text{in}}\!\!\\
\end{array}%
\right)\\
&+&\frac{\kappa}{\sqrt{T}}\int_{0}^tdt'\left(
                                                           \begin{array}{c}
                                                             \!\!\cos(\Omega t')\!\! \\
                                                             \!\!\sin(\Omega t')\!\!\\
                                                           \end{array}
                                                         \right)\bar{p}_{\text{L}}(ct',0),\nonumber
\end{eqnarray}
where $\tilde{\alpha}_{x/p}(t)$ describe the time dependence of $H_{\text{extra,II}}$
in the rotating frame (see Eq.~(\ref{Eq:RotationMatrix})). The resulting time evolution on Bob's side can be evaluated by inserting Eq.~(\ref{Eq:EvolutionCharlie}) into Eq.~(\ref{Eq:Teleportation}).
As in the general case above, the resulting differential equation can be approximated by the coarse-grained equation with $M^{\Omega,\Omega}_{g,-g}\rightarrow -\frac{1}{2}\openone$ for $\Omega \Delta t\gg 1$, where $\Delta t$ is the coarse graining time interval. This yields
\begin{eqnarray}\label{Eq:TimeDependentTeleportation}
\left(%
\begin{array}{c}
  \!\!\dot{\tilde{x}}_{\text{I}}^{\text{fin}}(t)\!\! \\
  \!\!\dot{\tilde{p}}_{\text{I}}^{\text{fin}}(t)\!\!\\
\end{array}%
\right)
\!\!&=&\!\!\frac{-\kappa}{2 T}\!\!\int_{0}^t \!\!\!\!dt'\left(
                                               \begin{array}{c}
                                                  \!\!\bar{g}_{\text{\tiny{x}}}\tilde{\alpha}_p(t')  \!\!\\
                                                 \!\! \bar{g}_{\text{\tiny{p}}}\tilde{\alpha}_x(t')  \!\!\\
                                               \end{array}
                                             \right)
\!\!+\!\!\frac{\kappa}{2 T}\!\left(%
\begin{array}{c}
  \!\!-\bar{g}_{\text{\tiny{x}}}[\tilde{x}_{\text{I}}^{\text{in}}\!+\!\tilde{x}_{\text{II}}^{\text{in}}]\!\! \\
  \!\!\bar{g}_{\text{\tiny{p}}}[\tilde{p}_{\text{I}}^{\text{in}}\!+\!\tilde{p}_{\text{II}}^{\text{in}}]\!\!\\
\end{array}%
\right)\nonumber\\
\!\!&+&\!\!\frac{ \kappa^2}{ T^{\frac{3}{2}}}\int_{0}^tdt'\left(
                                                           \begin{array}{c}
                                                             \!\!-\bar{g}_{\text{\tiny{x}}}\cos(\Omega t')\!\! \\
                                                             \!\!\bar{g}_{\text{\tiny{p}}}\sin(\Omega t')\!\!\\
                                                           \end{array}
                                                         \right)\bar{p}_{\text{L}}(ct',0)\nonumber\\
\!\!\!\!&+&\!\!\!\!\frac{1}{\sqrt{T}}\!\left(
             \begin{array}{cc}
              \!\!\bar{g}_{\text{\tiny{x}}} \sin(\Omega t)\! &\! \kappa\cos(\Omega t)\!\! \\
             \!\!\bar{g}_{\text{\tiny{p}}} \cos(\Omega t)\! & \!\kappa \sin(\Omega t) \!\!\\
             \end{array}
           \right)\!\!\left(
                   \begin{array}{c}
                    \!\! \bar{x}_{\text{L}}(ct,0) \!\!\\
                   \!\! \bar{p}_{\text{L}}(ct,0)\! \!\\
                   \end{array}
              \right)\!.
\end{eqnarray}
The first term on the right is equivalent to a field experienced by Bob. Bob's ensemble evolves as if it was placed in a magnetic field, whose time dependence is given by Charlie's evolution (see Eq.~(\ref{Eq:EvolutionCharlie})). This way, the effect of the magnetic field applied at Charlie's side is teleported to Bob.
This is possible, since entanglement generation, measurement and feedback are performed continually. In contrast to a traditional teleportation, Charlie's quantum state is therefore not destroyed in a single step and subsequently restored on Bob's side, but rather transmitted continuously, which offers the possibility to include the effect of a time evolution.

In the following, we discuss imperfections and consider the case, where real-time feedback is applied to both ensembles. This discussion can also be applied for evaluating the added noise in the teleportation protocol.
For analyzing the performance of the scheme, we introduce a figure of merit which quantifies the deviation of the realized time evolution from the desired one using the Jamiolkowsi isomorphism between quantum maps and states. Both, the imperfect map and the ideal one are transformed into their corresponding states and then compared as explained in the SM.
The time evolution $\varepsilon_{T}$ acting on the atomic system is described in terms of an entangled state of twice the system size that can be used to teleport a given input state $\rho_{\text{in}}$ through $\varepsilon_{T}$, such that the output $\varepsilon_{T}(\rho_{\text{in}})$ is obtained. This entangled state consists of two copies of a two mode squeezed state with squeezing parameter $R$. For $R\rightarrow\infty$, which corresponds to a quantum state with infinite energy, any input state can be teleported through $\varepsilon_{T}$. For finite $R$, this holds for a restricted set of input states.
We start by considering the fidelity for fixed $R$. As explained in the SM, the scheme involves two types of imperfections, fast rotating terms in the evolution and light noise added to the atomic system. If a time window $\Delta t\gg \Omega_1,\Omega_2,|\Omega_1-\Omega_2|$ is considered, the former are negligible and the latter can be suppressed using squeezed input light fields. The inset in Fig.~\ref{Fig:Plot1} shows the fidelity for fixed $R$ versus the squeezing of the light field $r$ for $\Omega_2=2\Omega_1$. For increasing $\Omega_1\Delta t$, the optimal squeezing parameter $r_{\text{opt}}$ increases, which leads to an increased accuracy of the scheme. For $\Omega_1\Delta t\rightarrow \infty$, $r_{\text{opt}}\rightarrow\infty$ and $F\rightarrow 1$.
For increasing values of $R$, correspondingly high values of $\Delta t$ are required to obtain a good fidelity. The required value of $\Delta t$ increases with $e^R$ as shown in Fig.~\ref{Fig:Plot1} which displays the fidelity $F$ for fixed Larmor frequencies $\Omega_1$ and $\Omega_2=2\Omega_1$. This graph features a fine substructure since for time windows $\Delta t=2\pi/\Omega_1$, local maxima are obtained, which gives rise to stripe-like regions of high fidelity.
In general, realizing a desired time evolution with high temporal resolution requires high Larmor frequencies. For fixed $\Omega_1$ and $\Omega_2$, the attainable precision depends on the temporal resolution as shown in Fig.~\ref{Fig:Plot1}. Higher fidelities can be obtained if a stroboscopic interaction is implemented where points of interest in time are chosen to coincide with the local maxima.

In conclusion, we proposed a dynamical teleportation scheme, where entangling operations, measurement and feedback are performed continuously and simultaneously. Moreover, we demonstrated how a generalized version of this protocol can be used to implement arbitrary quadratic interactions between remote systems.
\begin{acknowledgements}
We thank Karl Gerd Vollbrecht and G\'{e}za Giedke and acknowledge
support by the EU project MALICIA, the Alexander von Humboldt Foundation and the Spanish program TOQATA (FIS2008-00784). Additional support by the ERC grants INTERFACE and QUAGATUA, the Danish National ScienceFoundation Center QUANTOP and the DARPA program QUASAR are also acknowledged.
\end{acknowledgements}

%

%
\newpage
\appendix
\renewcommand{\figurename}{Supplementary figure}
\setcounter{figure}{0} \setcounter{equation}{0}
\renewcommand{\thefigure}{S.\arabic{figure}}
\renewcommand{\theequation}{S.\arabic{equation}}
\widetext

\section*{Supplemental Material}

In the following, we discuss imperfections of the proposed scheme for inducing nonlocal dynamics and show that the desired time evolution can be realized perfectly in the ideal case. In Sec.~\ref{SM:Imperfections}, we consider the different types of imperfection and explain in which parameter regime the desired time evolution can be faithfully implemented. In Sec.~\ref{SM:TimeEvolutions} we show how the deviation of the implemented time evolution from the desired one can be quantified and introduce an adequate figure of merit. In Sec.~\ref{SM:IncludeSqueezing}, we include squeezed light in the consideration and calculate the resulting fidelity of the atomic evolution numerically. We consider here the general scheme involving feedback on both ensembles for generating an effective interaction between two atomic ensembles. The noise contribution in the teleportation scheme can be analyzed along the same lines as the added noise discussed in the following (in particular in Sec.~\ref{SM:IncludeSqueezing}).
\subsection{Imperfections of the scheme}\label{SM:Imperfections}
Throughout the Supplemental Material, we consider two different time scales
\begin{eqnarray*}
\tau\ll \Omega^{-1}\ll\Delta t,
\end{eqnarray*}
and assume that the frequencies $\Omega_1$ ,$\Omega_2$, and $|\Omega_1-\Omega_2|$ take values of the same order of magnitude, which defines a time scale $\Omega^{-1}$. $\tau\ll \Omega^{-1}$ is an infinitesimally short time interval which will be used to express a continuous time evolution in a discretized form. $\Delta t \gg \Omega^{-1}$ is length of the considered time window.\\
\\We consider the setup shown in Fig.~1a in the main text for establishing an effective interaction between two atomic ensembles. The time evolution of the atomic system during an infinitesimal time step of duration $\tau$ is given by
\begin{eqnarray}\label{SM:Eq:InfinitesimalEvolution}
\left(
  \begin{array}{c}
    \tilde{x}_{\text{I}}([n+1]\tau) \\
    \tilde{p}_{\text{I}}([n+1]\tau) \\
    \tilde{x}_{\text{II}}([n+1]\tau) \\
    \tilde{p}_{\text{II}}([n+1]\tau) \\
  \end{array}
\right)&=&\left(\openone+G(n\tau)\ \!\tau\right)
\left(
  \begin{array}{c}
    \tilde{x}_{\text{I}}(n\tau) \\
    \tilde{p}_{\text{I}}(n\tau) \\
    \tilde{x}_{\text{II}}(n\tau) \\
    \tilde{p}_{\text{II}}(n\tau) \\
  \end{array}
\right)
+\left(
                      \begin{array}{c}
                        N_{x_1,n} \\
                        N_{p_1,n}  \\
                        N_{x_2,n}  \\
                        N_{p_2,n}  \\
                      \end{array}
                    \right).
\end{eqnarray}
The second term on the right is an undesired contribution due to the mapping of the input-quadratures of the light field onto the atomic ensembles,
\begin{eqnarray}\label{SM:Eq:NoiseInfinitesimal}
\left(
                      \begin{array}{c}
                        N_{x_1,n} \\
                        N_{p_1,n}  \\
                        N_{x_2,n}  \\
                        N_{p_2,n}  \\
                      \end{array}
                    \right)=\frac{\kappa }{\sqrt{N}}\left(
                      \begin{array}{c}
                         \cos(\Omega_1 t)\\
                         \sin(\Omega_1 t)\\
                         \cos(\Omega_2 t)\\
                         \sin(\Omega_2 t)\\
                      \end{array}
                    \right)p_{L,n}^{\text{in}}-\frac{g}{\sqrt{N}}\left(
                      \begin{array}{c}
                         \frac{1}{Z}\sin(\Omega_2 t)\\
                         Z \cos(\Omega_2 t)\\
                         Z \sin(\Omega_1 t)\\
                         \frac{1}{Z}\cos(\Omega_1 t)\\
                      \end{array}
                    \right)x_{L,n}^{\text{in}},
\end{eqnarray}
where $t=n\tau$. The matrix $G$ is given by
\begin{eqnarray*}
G(t)=\frac{g\kappa}{T}\left(
                          \begin{array}{cccc}
                            \frac{1}{Z} \sin(\Omega_2 t)\sin(\Omega_1 t) & -\frac{1}{Z}\sin(\Omega_2 t)\cos(\Omega_1 t) &\frac{1}{Z} \sin(\Omega_2 t) \sin(\Omega_2 t)& -\frac{1}{Z}\sin(\Omega_2 t)\cos(\Omega_2 t) \\
                            Z \cos(\Omega_2 t)\sin(\Omega_1 t)& -Z\cos(\Omega_2 t)\cos(\Omega_1 t) & Z\cos(\Omega_2 t)\sin(\Omega_2 t) & -Z\cos(\Omega_2 t) \cos(\Omega_2 t)\\
                            Z \sin(\Omega_1 t)\sin(\Omega_1 t)& -Z\sin(\Omega_1 t)\cos(\Omega_1 t) & Z\sin(\Omega_1 t) \sin(\Omega_2 t)& -Z\sin(\Omega_1 t)\cos(\Omega_2 t) \\
                            \frac{1}{Z}\cos(\Omega_1 t)\sin(\Omega_1 t) & -\frac{1}{Z}\cos(\Omega_1 t)\cos(\Omega_1 t) & \frac{1}{Z}\cos(\Omega_1 t) \sin(\Omega_1 t)&-\frac{1}{Z} \cos(\Omega_1 t)\cos(\Omega_2 t) \\
                          \end{array}
                        \right).
\end{eqnarray*}
The dynamics described by Eq.~(\ref{SM:Eq:InfinitesimalEvolution}) has to be compared to the desired time evolution for the atomic system,
\begin{eqnarray*}
\dot{\vec{X}}(t)=G_0 \vec{X}(t)
\end{eqnarray*}
where $\vec{X}=(\tilde{x}_{\text{I}} , \tilde{p}_{\text{I}} , \tilde{x}_{\text{II}} , \tilde{p}_{\text{II}} )^T$ and
\begin{eqnarray*}
G_0=\frac{g\kappa}{2T}\left(
                          \begin{array}{cccc}
                            0 & 0 & \frac{1}{Z} & 0 \\
                            0 & 0 & 0 & -Z\\
                            Z & 0 & 0 &0\\
                            0 & -\frac{1}{Z} & 0 & 0 \\
                          \end{array}
                        \right).
\end{eqnarray*}
The scheme involves two types of imperfections. Firstly, the proposed scheme does not implement exactly the atomic time evolution matrix $G_{0}$ but rather the time-dependent matrix $G(t)$. Secondly, light noise is added to the system (see Eq.~(\ref{SM:Eq:NoiseInfinitesimal})). We start by discussing the first type of imperfection. To this end, we are considering the time evolution during a time window $\Delta t$ in the absence of added noise. The solution of the differential equation $\dot{\vec{X}}(t)=G(t) \vec{X}(t)$ can be approximated by $\vec{X}(t)=e^{G_0 t}\vec{X}(0)$ in the limit
\begin{eqnarray*}
\epsilon_{1,1}&=&(\Omega_1\Delta t)^{-1}\ll 1,\\
\epsilon_{2,2}&=&(\Omega_2\Delta t)^{-1}\ll 1,\\
\epsilon_{1,2}&=&(|\Omega_1-\Omega_2|\Delta t)^{-1}\ll 1.
\end{eqnarray*}
This is due to the fact that the matrix elements involving two different trigonometric functions average out upon integration, for example
\begin{eqnarray*}
\frac{1}{\Delta t}\int_t^{t+\Delta t}dt'\sin(\Omega_i t')\cos(\Omega_i t')&=&\mathcal{O}(\epsilon_{i,i}),\ \ \ \ \
\frac{1}{\Delta t}\int_t^{t+\Delta t}dt'\sin(\Omega_i t')\sin(\Omega_j t')=\mathcal{O}(\epsilon_{i,j}), \ \text{for}\  i \neq j,
\end{eqnarray*}
while
\begin{eqnarray*}
\frac{1}{\Delta t}\int_t^{t+\Delta t}dt'\sin^2(\Omega_{i}t')&=&\frac{1}{2}+\mathcal{O}(\epsilon_{i,i}),\ \ \ \ \ \ \
\frac{1}{\Delta t}\int_t^{t+\Delta t}dt'\cos^2(\Omega_{i}t')=\frac{1}{2}+\mathcal{O}(\epsilon_{i,i}).
\end{eqnarray*}
The effective time evolution is valid on a coarse-grained time scale $\Delta t\gg \Omega_1^{-1},\Omega_1^{-2},|\Omega_1-\Omega_2|^{-1}$. This can be understood in analogy to familiar examples of effective $2^{\text{nd}}$ order Hamiltonians such as AC stark shift Hamiltonians.\\
\\As a next step, we include the noise terms in the discussion and consider the complete time evolution $\dot{\vec{X}}(t)=G(t) \vec{X}(t)+\vec{N}(t)$ and its solution
\begin{eqnarray}\label{SM:Eq:IntegratedSolution}
\vec{X}(t)=\mathcal{T}\ \!e^{\int_{0}^{ t}G(\tau)d\tau}\ \!\vec{X}(0)+\vec{N}^{\text{int}},
\end{eqnarray}
where $\mathcal{T}$ is the time ordering operator and
\begin{eqnarray*}
\vec{N}^{\text{int}}=\int_0^{t}\mathcal{T}     e^{\int_{t'}^{t}G(\tau')d\tau'}\ \!\vec{N}(t')dt'.
\end{eqnarray*}
The components of the integrated noise operator $\vec{N}^{\text{int}}=\left(
                                                       \begin{array}{cccc}
                                                         N_{x1}^{\text{int}} & N_{p1}^{\text{int}} & N_{x2}^{\text{int}} & N_{p2}^{\text{int}} \\
                                                       \end{array}
                                                     \right)^T
$
are mutually independent
\begin{eqnarray*}
[N^{\text{int}}_{i},N^{\text{int}}_{j}]=0,
\end{eqnarray*}
for $\epsilon\rightarrow 0$, where $\epsilon=\text{max}(\epsilon_{1,1},\epsilon_{2,2},\epsilon_{1,2})$. Therefore, all four noise contributions can be squeezed simultaneously in this limit. Note that the two types of imperfections that occur in this scheme are intimately linked. For $\epsilon\rightarrow 0$, the first part of the differential equation corresponds to a unitary time evolution, which implies that the noise terms can be squeezed such that their contribution to the dynamics becomes negligible.
\subsection{Comparison of time evolutions}\label{SM:TimeEvolutions}
In order to asses the performance of the scheme, we need to quantify how close the established time evolution is to the desired one. We do this in terms of covariance matrices using the Jamiolkowski isomorphism as explained in Sec.~\ref{SM:Sec:Jamiolkowski}. The corresponding figure of merit is introduced in Sec.~\ref{SM:Sec:FigureOfMerit}.
\subsubsection{Characterization of linear time evolutions using The Jamiolkowski isomorphism}\label{SM:Sec:Jamiolkowski}
\begin{figure}
 \includegraphics[width=0.4\columnwidth]{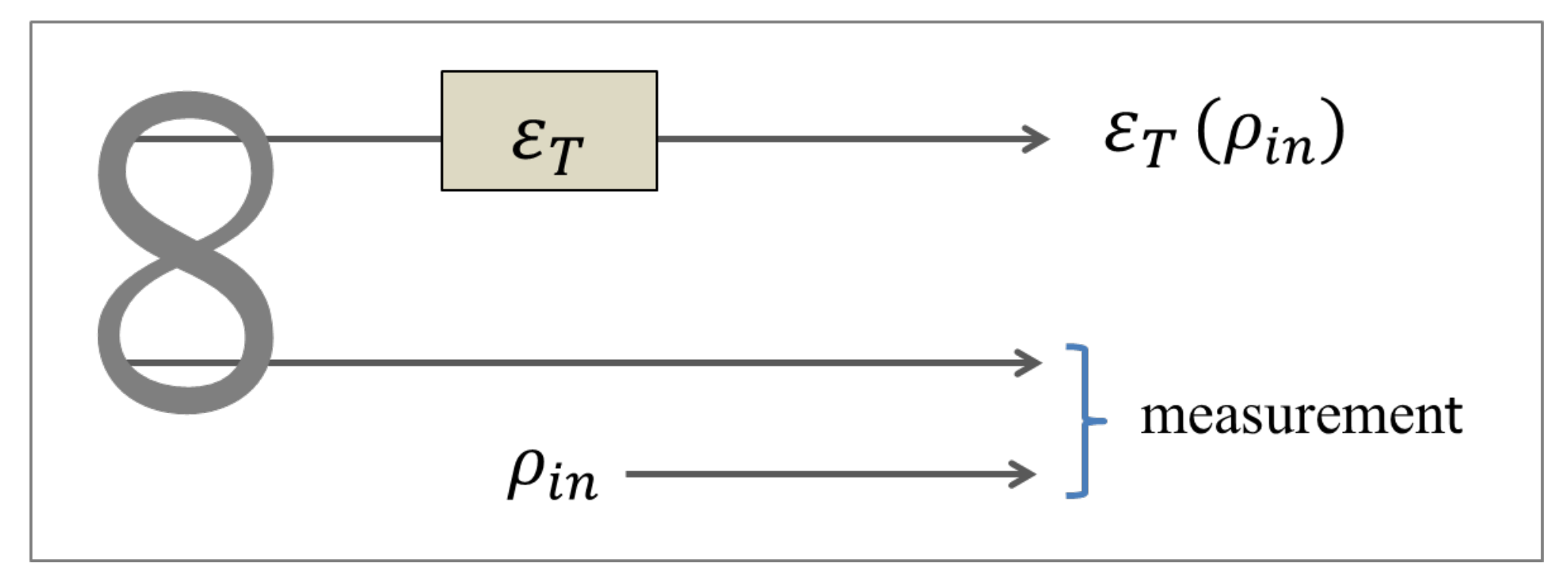}
 \caption{Characterization of a completely positive map $\epsilon_{T}$ in terms of a quantum state using the Jamiolkowski isomorphism. An operation $\epsilon_{T}$ acting on quantum system can be represented by the density matrix $\rho_{\text{J}}$, which corresponds to twice the system size. $\rho_{\text{J}}$  can be interpreted as two entangled copies of a quantum state, where $\epsilon_{T}$ has been applied to one subsystem. Given the state $\rho_{\text{J}}$, the map $\epsilon_{T}$ can be (probabilistically) implemented by teleporting an input state $\rho_{in}$ through $\epsilon_{T}$. This can be done by performing a suitable joined measurement on $\rho_{in}$ and one subsystem.}
 \label{SM:Fig:Jamiolkowski}
 \end{figure}

The Jamiolkowski isomorphism allows one to describe a time evolution in terms of states.
Gaussian quantum states, i.e. states with a Gaussian Wigner function, are completely characterized by their first and second moments. They can be conveniently described in terms of their displacement vector $D$, with components $D_i=\langle X_i\rangle$, and their covariance matrix $\Gamma$, with $\Gamma_{ij}=\langle\{X_i,X_j\}_+ \rangle$, where $\{\cdot,\cdot\}_+$ is the anticommutator \cite{GaussianStates}.
In the following, we use the Jamiolkowski isomorphism in order to characterize a time evolution of the atomic system in terms of a quantum state $(\Gamma,D)$, as explained in~\cite{Giedke2002}. To this end, we consider two identical two mode squeezed states with squeezing parameter $R$. The corresponding covariance matrix $\Gamma_{\text{J}}$ for the two entangled copies of the atomic system is given by
\begin{eqnarray*}
\Gamma_{\text{J}}&=&\left(
                      \begin{array}{cc}
                        A(R) & C(R) \\
                        C(R)^T & A(R) \\
                      \end{array}
                    \right),
\end{eqnarray*}
where $A(R)$ and $C(R)$ are $4\times4$ matrices,
\begin{eqnarray*}
A(R)=\cosh(R)\openone, \ \ \ \ \ C(R)=\sinh(R) \text{diag}(1,-1,1,-1).
\end{eqnarray*}
In the limit $R\rightarrow\infty$, the two mode squeezed state is the improper maximally entangled state $|\Psi_{\text{TMSS,}\infty}\rangle\propto \sum_{n=0}^{\infty}|n,n\rangle$.
A time evolution of the atomic system $\vec{X}(t)= S(t)\vec{X}(0)+n(t)$ (where $S(t)$ is a  time evolution matrix and $n(t)$ added noise) can be characterized by applying the corresponding completely positive map $\varepsilon_{T}$ to one part of the entangled composite quantum state introduced above, which results in the covariance matrix
\begin{eqnarray*}
\Gamma_{\text{J}[S,n]}&=&\left(
                      \begin{array}{cc}
                        SA(R)S^T+\Gamma_{\text{\tiny{noise}}} & SC(R) \\
                        (SC(R))^T & A(R) \\
                      \end{array}
                    \right),
\end{eqnarray*}
where $\Gamma_{\text{\tiny{noise}}}=nn^T+n^Tn$. Fig.~\ref{SM:Fig:Jamiolkowski} illustrates an intuitive explanation of this characterization. The time evolution $\varepsilon_{T}$ can be applied to an input state $\rho_{\text{in}}$ by teleporting this state by means of a joint measurement with one part of the composite entangled system (see~\cite{Giedke2002} for a more detailed explanation).
For $R\rightarrow\infty$, any input state $\rho_{\text{in}}$ can be probabilistically teleported in this way, resulting in the output $\varepsilon_{T}(\rho_{\text{in}})$. For finite values of $R$, this reasoning can be applied to a restricted set of input states.\\
\\A noisefree time evolution $\varepsilon_{T}$ translates into a Jamiolkowski covariance matrix $\Gamma_{\text{J}}$ that corresponds to a pure state. Deviations of the implemented time evolution from the desired unitary one translate directly into noise added to the Jamiolkowski matrix.
\subsubsection{Figure of merit}\label{SM:Sec:FigureOfMerit}
In order to assess the deviation of the covariance matrix that corresponds to the time evolution implemented by the proposed scheme $\Gamma_{\Delta t}$,  from the ideal (desired) one, $\Gamma_{\infty}$, we introduce the figure of merit
\begin{eqnarray*}
E&=&\text{max}_{\vec{C}}\left[\frac{\vec{C}^{\dag}\left(\Gamma_{\Delta t}-\Gamma_{\infty}\right)\vec{C}}{\vec{C}^{\dag}\Gamma_{\Delta t}\vec{C}}\right],
\end{eqnarray*}
which quantifies the error. Maximization over all complex vectors $\vec{C}$ yields
\begin{eqnarray}\label{EqSM:NoiseFormula}
E = 1-m({\Gamma_{E}}),
\end{eqnarray}
where $m({\Gamma_{E}})$ is the minimum eigenvalue of the matrix
\begin{eqnarray}\label{EqSM:Gamma}
\Gamma_{E}=\frac{1}{\sqrt{\Gamma_{\Delta t}}}\ \!\Gamma_{\infty}\frac{1}{\sqrt{\Gamma_{\Delta t}}}.
\end{eqnarray}
This can be seen by noting that
\begin{eqnarray*}
E=1-\text{min}_{\vec{C}}\left[\frac{\vec{C}^{\dag}\Gamma_{\infty}\vec{C}}{\vec{C}^{\dag}\Gamma_{\Delta t}\vec{C}}\right],
\end{eqnarray*}
and considering the derivative of the expression $\lambda_{\infty}/\lambda_{\Delta t}$ with respect to $\vec{C}$, where
\begin{eqnarray*}
\lambda_{\Delta t}=\vec{C}^{\dag}\Gamma_{\Delta t}\vec{C},\ \ \ \ \ \
\lambda_{\infty}=\vec{C}^{\dag}\Gamma_{\infty}\vec{C}.
\end{eqnarray*}
The resulting condition for extremality
\begin{eqnarray*}
\lambda_{\Delta t}\Gamma_{\infty}\vec{C}-\lambda_{\infty}\Gamma_{\Delta t}\vec{C}=0
\end{eqnarray*}
translates directly into
\begin{eqnarray*}
\frac{\lambda_{\infty}}{\lambda_{\Delta t}}\vec{d}=\Gamma_{E} \vec{d},
\end{eqnarray*}
where $\Gamma_{E}$ is given by Eq.~(\ref{EqSM:Gamma}) and $\vec{d}=\sqrt{\Gamma_{\Delta t}} \vec{C}$.
\subsection{Fidelity of the scheme for squeezed input states}\label{SM:IncludeSqueezing}
\begin{figure}
 \includegraphics[width=0.6\columnwidth]{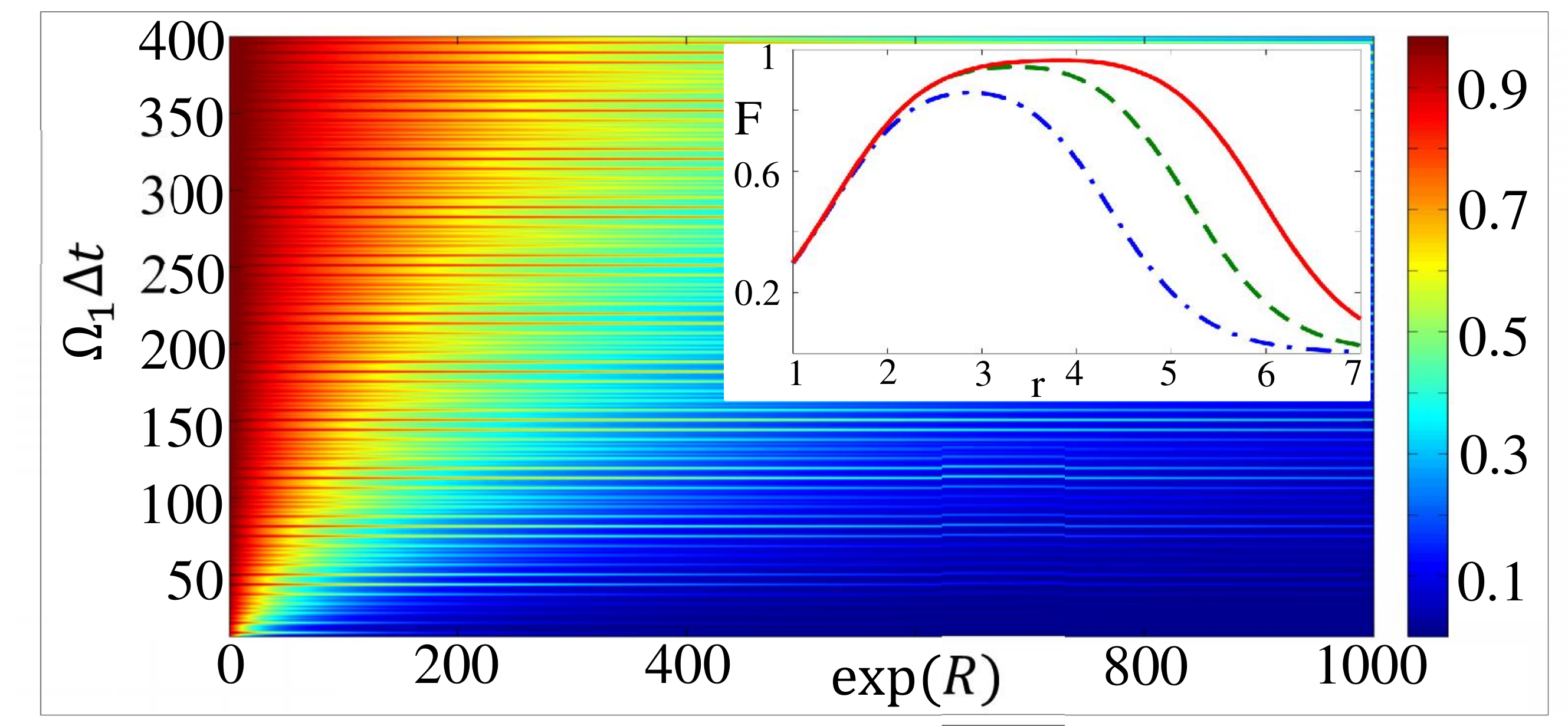}
 \caption{Fidelity for for realizing a beamsplitter interaction for fixed Larmor frequencies $\Omega_1$ and $\Omega_2=2\Omega_1$. The feedback-gain is chosen such that the resulting interaction strength corresponds to the underlying QND interaction $g=\kappa=1$. The corresponding results for implementing a two mode squeezing Hamiltonian are shown in Fig.~2 in the main text. The main panel displays the fidelity $F=1-E$ for optimal squeezing parameter $r_{\text{opt}}$ of the input light field versus $e^R$ and $\Omega_1\Delta t$. The inset shows the attainable fidelity versus the squeezing parameter $r$ of the input light field for $R=3$. The three curves correspond in ascending order to $\Omega_1\Delta t=100$ (blue dash-dotted line), $\Omega_1\Delta t=300$ (dashed green line) and $\Omega_1\Delta t=500$ (solid red line).}
 \label{Fig:SM:Plot2}
 \end{figure}

In this section, we show that the desired time evolution can be implemented perfectly if squeezed light fields are used. To this end, we explain how squeezed photonic input states can be included in the derivation and calculate the attainable fidelities.\\
\\The time evolution of the atomic system is given by Eq.~(\ref{SM:Eq:IntegratedSolution}). The total (integrated) noise contribution $N^{\text{int}}=(N_{x1}^{\text{int}}, N_{p1}^{\text{int}}, N_{x2}^{\text{int}}, N_{p2}^{\text{int}})^T$
can be written in the form
\begin{eqnarray*}
\vec{N}^{\text{int}}=T \ \!\vec{Y},
\end{eqnarray*}
where $\vec{Y}=(x_{\text{L,}1},x_{\text{L,}2},...,x_{\text{L,}N},p_{\text{L,}1},p_{\text{L,}2},...,p_{\text{L,}N})^T$ is the input light field vector and $T$ is a $4\times 2N$ matrix. We can define canonical partners $\vec{N'}^{\text{int}}=({N'}_{x1}^{\text{int}}, {N'}_{p1}^{\text{int}}, {N'}_{x2}^{\text{int}}, {N'}_{p2}^{\text{int}})^T$ and normalize the operators such that
\begin{eqnarray*}
[N^{\text{int}}_{i},{N'}^{\text{int}}_{j}]&=&i\ \!\delta_{i,j}, \ \ \ \ \ i,j=1,2,3,4.
\end{eqnarray*}
Using these definitions, we can define the vector $\vec{N}^{\text{int}}_{8}=\left(N_{x1}^{\text{int}}, N_{p1}^{\text{int}}, N_{x2}^{\text{int}}, N_{p2}^{\text{int}},{N'}_{x1}^{\text{int}}, {N'}_{p1}^{\text{int}}, {N'}_{x2}^{\text{int}}, {N'}_{p2}^{\text{int}}\right)^T$, where the subscript indicates that this vector includes the canonical partners. The vector $\vec{N}^{\text{int}}_8$ can be written in matrix form
\begin{eqnarray*}
\vec{N}^{\text{int}}_8=T_8 \ \!\vec{Y}.
\end{eqnarray*}
The noise operators contribution to the atomic evolution $N_{x1}^{\text{int}}$, $N_{p1}^{\text{int}}$, $N_{x2}^{\text{int}}$, $N_{p2}^{\text{int}}$ are independent for $\epsilon\rightarrow 0$. In this limit, the total noise commutation relation is given by
$
[N^{\text{int}}_{8,i},{N}^{\text{int}}_{8,j}]=i\ \!(\sigma_{\infty,4})_{ i,j}
$
with
\begin{eqnarray*}
\sigma_{\infty,n}=\left(
                  \begin{array}{cc}
                    0_{n} & \openone_n \\
                   - \openone_n & 0_{n} \\
                  \end{array}
                \right),
\end{eqnarray*}
where $0_{n}$ and $\openone_n$ are the $n\times n$ zero and unit matrix respectively. For finite values of $\epsilon$,
\begin{eqnarray*}
\sigma_{\Delta t}=T_8\ \!\sigma_{\infty,2N}\ T_8^T=\sigma_{\infty,4}+O(\epsilon).
\end{eqnarray*}
Due to the small correction on the right side, the noise modes can not be squeezed simultaneously. We consider here a squeezing operation with squeezing parameter $r$  on the input light field which is applied prior to the actual protocol. More specifically, we consider the operation on the input light field that corresponds to a squeezing of the integrated noise modes~\cite{FootnoteSqueezing}. The fact that $\sigma_{\Delta t}\neq\sigma_{\infty,4}$ leads to a tradeoff that prevents the fidelity of the scheme to grow continuously with increasing squeezing parameter $r$. Therefore, there is an optimal squeezing parameter $r_{\text{opt}}$ for a given value of $\epsilon$. This can be understood by expressing the integrated noise modes in terms of canonically commuting modes $\vec{n}_8=(n_1,n_2,n_3,n_4,n'_1,n'_2,n'_3,n'_4)^T$ (with $[n_i,n_j]=i (\sigma_\infty,4)_{ij}$), such that the noise contribution to the atomic covariance matrix can be written as
\begin{eqnarray*}
\Gamma_{\text{noise}}=\alpha \ \!\Gamma_n +\beta \ \!\Gamma_{n'} \ ,
\end{eqnarray*}
with real coefficients $\alpha$ and $\beta$, where $\beta$ is on the order of $\epsilon$. $\Gamma_n$ and $\Gamma_{n'}$ are the covariance matrices corresponding to $\vec{n}=(n_1,n_2,n_3,n_4)^T$ and $\vec{n}'=(n'_1,n'_2,n'_3,n'_4)^T$ respectively. Considering a squeezed state of the modes $\vec{n}$ with $\Gamma_n=e^{-r}\openone_4$ and $\Gamma_{n'}=e^{r}\openone_4$ leads to the tradeoff described above. The insets of Fig.~\ref{Fig:SM:Plot2} and Fig.~2 in the main text show the fidelity $F=1-E$ (where $E$ is the error as defined in Sec.~\ref{SM:Sec:FigureOfMerit}) versus the input squeezing $r$ for fixed values of $\Omega_1$, $\Omega_2$  and $R$. We analyze here the realization of an active (two mode squeezing) and a passive (beamsplitter like) interaction, since any evolution under a quadratic Hamiltonian can be realized by combining these two interactions with local rotations~\cite{HamiltonianSimulation,Bennett2002}.
The plot shows how decreasing values of $\epsilon$ lead to an increased optimal squeezing parameter $r_{\text{opt}}$ and therefore to an increased accuracy of the scheme.\\
For high values of the parameter $R$, correspondingly high values of $\Delta t$ are required to obtain a good fidelity according to the definition given in Sec.\ref{SM:Sec:FigureOfMerit}. More specifically, the optimal fidelity depends on the product $e^R\Delta t^{-1}$ as shown in Fig.~\ref{Fig:SM:Plot2} which depicts $F=1-E$ versus $e^R$ and $\Delta t$ for fixed Larmor frequencies $\Omega_1$ and $\Omega_2=2\Omega_1$. Apart from that, the graph features a fine substructure of local maxima for $\Delta t=2\pi/\Omega_1$.\\
Fig.~\ref{Fig:SM:Plot2} and Fig. 2 in the main text depict the fidelity for the minimum interaction time $T=\Delta t$. Using longer interaction times $T=f \Delta t$ (with the same error $E$) corresponds to increasing the values of $\kappa$ and $g$ by a factor of $\sqrt{f}$~\cite{FootnoteKappa}.\\
\\In summary, high fidelities can be obtained for $\epsilon \ll 1$, which requires $\Delta t\gg \Omega_1,\Omega_2,|\Omega_1-\Omega_2|$. As outlined in the main text, this relation can be understood in terms of a time coarse graining with a coarse graining time interval $\Delta t$. The realization of a desired time evolution with a very good temporal resolution requires high Larmor frequencies. For fixed Larmor frequencies $\Omega_1$ and $\Omega_2$, the attainable precision depends on the temporal resolution as shown in the figures. We remark that higher fidelities can be obtained if one realizes a stroboscopic interaction where points of interest in time are chosen to coincide with the phase matching condition.

\begin{thebibliography}{54}
\bibitem{FootnoteTeleportation}
Because of the inherent fragility of quantum information, the direct (physical) transport is difficult to achieve with high fidelity. Moreover, the no-cloning theorem~\cite{noCloning} poses fundamental limitations on strategies based on measurements and re-preparation.
\bibitem{FootnoteEnsembles}
The spin state of strongly polarized ensembles can be
described by bosonic operators within the Holstein Primakoff
approximation~\cite{HP40,HSP10}.
\bibitem{FootnoteNeglectR}
We neglect the distance $R$ between the ensembles, since ut is assumed that the time delay
$\delta t=R/c$ is short on the time scale of atomic transitions $1/\Gamma_{\text{eff}}$, such that retardation effects are negligible. For the setup under consideration~\cite{HSP10}, where $R$ is on the order of a meter and $\Gamma_{\text{eff}}\ll \Gamma$ (where $\Gamma$ is the natural linewidth of the involved excited states), this condition is fullfilled. In our protocols, non-negligible retardation effects would lead to differential equations which are nonlocal in time.
\bibitem{FootnoteRotation}
We assume here that the rotation of the atomic system due to the
magnetic field is negligible during the short time interval $\tau$
(compare~\cite{HPC05}).
\bibitem{FootnoteHamiltonian1}
We consider here only interaction terms. Local terms can be either or passive or of active type. The former can be realized by applying magnetic fields. The latter can for example be realized by combining a QND interaction with continuous measurements on the light and real-time feedback.
\bibitem{FootnoteHamiltonian2}
The two mode squeezing Hamiltoninan $H_{\text{A}}$ is given by
$H_{\text{A}}\propto i\left(a_{1}a_{2}-a^{\dag}_{1}a^{\dag}_{2}\right)$ where
$a_{j}=\frac{1}{\sqrt{2}}\left(x_{j}+ip_{j}\right)$, with $j=1,2$.
$H_{\text{P}}\propto i\left(a_{1}a_{2}^{\dag}-a_{1}^{\dag}a_{2}\right)$ describes a beamsplitter-like interaction.
\bibitem{FootnoteLightmodes}
The continuous light modes with quadratures $\bar{x}_{\text{L}}(\xi,t)$, $\bar{p}_{\text{L}}(\xi,t)$ are obtained by considering spatially localized modes ${x}_{\text{L}}(z,t)$, ${p}_{\text{L}}(z,t)$ with $[{x}_{\text{L}}(z,t),{x}_{\text{L}}(z',t)]=ic\delta(z-z')$ and performing the variable transformation $\xi=ct-z$ (such that $\bar{x}_{\text{L}}(\xi,t)=x_{\text{L}}(ct-\xi,t)$).
\end{thebibliography}
\begin{thebibliography}{54}
\bibitem[S5]{FootnoteSqueezing}
The operation on the input light field that corresponds to a squeezing operation on the integrated noise mode can be described in terms of the Hamiltonian $H=(\vec{N}_8^{\text{int}})^T M_8 \vec{N}_8^{\text{int}}$, where $M_8=-\frac{r}{2\ \!T_{\text{squ}}}\left(
                                                                                \begin{array}{cc}
                                                                                  0_4 & \openone_4 \\
                                                                                  \openone_4 & 0_4 \\
                                                                                \end{array}
                                                                              \right)
$. $T_{\text{squ}}$ is the time period during which the squeezing Hamiltonian is applied. This leads to the differential equation $\dot{\vec{N}}_8^{\text{int}}=\mathcal{G}_8\vec{N}_8^{\text{int}}$ with $\mathcal{G}_8=\sigma_{\Delta t}(M_8+M_8^T)$.
\bibitem[S6]{FootnoteKappa} For atomic ensembles, the QND-coupling strength $\kappa$ is proportional to the square root of the number of atoms in the ensemble $\sqrt{N_{\text{A}}}$ and the square root number of photons in the light pulse $\sqrt{N_{\text{Ph}}}$, $\kappa\propto \sqrt{N_{\text{A}}N_{\text{Ph}}}$. Increasing the pulse duration $T$ by a factor of $f$ (using the same constant photon flux) corresponds therefore to an increase of the QND coupling strength $\kappa$ by a factor $\sqrt{f}$ .
\end{thebibliography}
\end{document}